\documentclass[twocolumn]{openjournal}

\usepackage{xcolor}
\usepackage{textgreek}
\usepackage[utf8]{inputenc}
\usepackage[english]{babel}

\usepackage{threeparttable}
\usepackage{booktabs}
\usepackage{rotating}
\usepackage{placeins}

\usepackage{color,colortbl}
\definecolor{linkcolor}{rgb}{0.0,0.3,0.5}
\usepackage{tensind}
\tensordelimiter{?}
\DeclareGraphicsExtensions{.bmp,.png,.jpg,.pdf}
\usepackage{verbatim}
\usepackage[normalem]{ulem}
\usepackage{orcidlink}
\usepackage{soul}

\graphicspath{ {./figs/} }
\usepackage{hyperref}
\hypersetup{
    unicode, 
    colorlinks=true,
    linkcolor=linkcolor,
    citecolor=linkcolor,
    filecolor=linkcolor,
    urlcolor=linkcolor,
}
\begin{document}
\title{A useful representation of TESS light curves}

\author{Dovi Poznanski \orcidlink{0000-0003-1470-7173}}
\email{dovi@tau.ac.il}
\affiliation{School of Physics and Astronomy, Tel-Aviv University Tel-Aviv 69978, Israel. \\}
\affiliation{Cahill Center for Astrophysics, California Institute of Technology, Pasadena CA 91125, USA.\\}
\affiliation{Kavli Institute for Particle Astrophysics \& Cosmology, 452 Lomita Mall, Stanford University, Stanford, CA 94305, USA. \\}
\affiliation{Department of Physics, Stanford University, 382 Via Pueblo Mall, Stanford, CA 94305, USA. \\}


\begin{abstract}
We present a simple and interpretable representation of TESS light curves designed for large-scale exploratory analysis. Our goal is not to optimize classification performance, but to construct a computationally efficient mapping in which proximity reflects meaningful similarity, without using labels or explicit period information as inputs. We represent each light curve using either quantile graphs or scattering transforms, reduce dimensionality with principal component analysis, and project the resulting features onto a self-organizing map (SOM). We evaluate $\sim$1500 model configurations using a combination of standard embedding diagnostics and a light-curve-shape-based cohesion metric, and select a compact quantile-graph-based model that balances interpretability, stability, and performance. Applying the model to $\sim$1.5 million TESS 2-minute cadence light curves, we find that the map organizes sources primarily by variability amplitude, signal-to-noise ratio, characteristic timescale, and light-curve shape. Repeat observations of the same stars show that most sources occupy stable and contiguous regions of the map, indicating that the representation captures persistent properties rather than noise and systematics. We provide an \href{http://tess-l8.space}{interactive web interface} that enables inspection of nodes, nearest neighbors, and individual sources across sectors. The resulting representation serves as a practical tool for exploration, anomaly detection, and dataset characterization, and illustrates how simple, deterministic encodings can yield useful structure in large astronomical time-series datasets.
\end{abstract}


\maketitle

\section{Introduction}
\label{s:intro}

Modern time-domain astronomy is no longer limited by the acquisition of light curves, but by our ability to organize, compare, and interpret them at scale. Surveys such as the Transiting Exoplanet Survey Satellite \citep[TESS; ][]{ricker15} now provide high-cadence photometry for hundreds of thousands of targets in short-cadence modes and millions more in full-frame products, while ground-based surveys such as the Zwicky Transient Facility \citep[ZTF; ][]{bellm19} and, imminently, the Rubin Legacy Survey of Space and Time \citep[LSST; ][]{ivezic19}, push the field further into the regime where many traditional approaches are no longer viable. In this setting, a central methodological question is how to represent variability in a way that is computationally feasible, scientifically useful, robust, and flexible.

Historically, most variable-star pipelines have relied on explicit feature engineering. Light curves are reduced to periods, amplitudes, colors, summary statistics, Fourier coefficients, or other hand-crafted descriptors, which are then used for classification or other tasks \citep[e.g.,][]{debosscher07,richards11,barbara22,gao25}. This approach has obvious strengths. It is interpretable, modular, and often physically well motivated. It also scales well. But it imposes a strong prior on what aspects of variability are considered important. As a result, it can miss structure that is not well captured by predefined features. Many such approaches also rely on period estimation, which can become computationally expensive at survey scale, despite recent progress \citep[e.g.,][]{finkbeiner25,boyle26}.

Many recent efforts explore learned or data-driven representations of light curves. Recurrent neural networks, transformers, and self-supervised methods have been used to learn embeddings that support classification, transfer learning, anomaly detection, or retrieval without requiring a fixed set of manually designed descriptors \citep[e.g., ][]{naul18,becker20,villar21,donoso-oliva23,zabriskie23,rizhko24,donoso-oliva25}. Data science in general, and the field of stellar variability in particular, are moving toward richer multimodal and foundation-style representations \citep[e.g.,][]{audenaert25}.

These developments are motivated not only by classification accuracy, but by the need for representations that can support a broader range of tasks. A useful representation can enable similarity search, outlier discovery, rapid visual inspection, transfer to downstream supervised models, and anomaly detection in large and heterogeneous samples. This is especially relevant in astronomy, where labels are sparse, taxonomies are incomplete, and the most interesting sources are often the unusual ones. Recent work has therefore increasingly emphasized unsupervised exploration and anomaly-oriented analysis of light curves, including on TESS data \citep[e.g.,][]{crake23}. In that setting, the goal is not merely to assign a class label, but to expose structure in the data in a way that remains scientifically interpretable.

TESS provides a particularly useful setting for this problem. Its 2-minute cadence products contain a very large sample of relatively high-quality light curves, while the full-frame images extend to vastly larger and more heterogeneous populations. Another useful aspect is the fact that stars are observed in multiple sectors, with a median of 3 visits, and a tail extending to 40, providing a natural baseline for similarity measures. Even within the 2-minute sample, the diversity of source types, variability amplitudes, signal-to-noise ratios, and observing baselines makes it difficult to define a single representation that is both efficient and broadly informative. A practical method must therefore tolerate heterogeneity, remain fast to evaluate, and still preserve enough structure to support human-guided exploration.

This paper takes a deliberately narrow approach. We do not attempt to build a universal latent space for all astronomical time series, nor a supervised classifier for a predefined taxonomy. Instead, we ask a simpler question: can one map raw, unfolded TESS light curves into a low-dimensional, interpretable space in which proximity usually corresponds to genuine light-curve similarity, without using labels or explicit period information as inputs? We are interested in a representation that is useful and fast before any downstream task is specified, and in understanding which design choices affect it. Such a representation could support exploratory analysis, quality control, neighborhood retrieval, anomaly detection \citep[see ][]{crake23}, and the construction of training sets for later supervised work.

To this end, we experiment with two families of representations, quantile-based summaries and scattering transforms, with dimensional compression and projection onto a self-organizing map (SOM). SOMs provide a discrete and topology-preserving organization of the data, allowing individual regions of the map to be inspected and interpreted directly \citep{kohonen95,armstrong16}. This property makes them particularly suitable for exploratory analysis, where human interpretation remains an essential component.

We apply the resulting pipeline to the full TESS 2-minute cadence sample that passes our quality cuts, about 95\% of the 1.5 million light curves, and use the learned map both as a scientific summary of the dataset and as the basis for an \href{http://tess-l8.space}{interactive web tool}. The web interface is intended to make the representation useful in practice: it allows users to inspect nodes, browse representative and outlying light curves, follow individual stars across sectors, and examine local neighborhoods on the map. In that sense, this work is not only about constructing a representation, but also about making a very large light-curve sample navigable.

The resulting map is not intended to replace existing classification catalogs or period-finding pipelines, nor is it intended to provide an optimal or universal representation of stellar time series. Instead, our goal is to explore the construction of a simple, transparent, and computationally efficient mapping that preserves meaningful structure in TESS light curves and supports human interpretation. Throughout the paper we therefore emphasize the trade-offs and impact of design choices, rather than maximizing performance on any single metric.

In the following sections, we describe the data and preprocessing (\S\ref{s:data}), the representation methods and SOM framework (\S\ref{s:methods}), the model selection procedure (\S\ref{s:model_select}), and the structure and application of the selected model to the full TESS 2-minute sample (\S\ref{s:best}--\S\ref{s:2mn}). We discuss our results and conclude (\S\ref{s:conclusions}), after a description of the interactive tool built on top of this representation (\S\ref{s:webtool}).

\section{Data and Pre-processing}
\label{s:data}
\subsection{TESS Light Curves}

We restrict ourselves to the 2-minute cadence TESS light curves, obtained from the Mikulski Archive for Space Telescopes(MAST). Specifically, the photometry processed by the Science Processing Operations Center (SPOC) pipeline \citet{jenkins16}, known as Presearch Data Conditioning Simple Aperture Photometry (PDCSAP; \citealt{smith12}; \citealt{stumpe12}; \citealt{stumpe14}) light curves. The PDCSAP light curves are based on an optimal aperture extraction of the raw photometry that was detrended to correct for instrument systematics. 

Mapping methods generally assume evenly sampled inputs. TESS light curves come close but never fully meet that requirement. We therefore impose a minimal continuity criterion. For each target–sector pair we search the light curve for a 4096-long contiguous segment, where we define continuous  as having no more than 150 missing observations in total, and no individual gap exceeding 30 minutes. We then treat these segments as continuous in subsequent steps, essentially ignoring the gaps. Downstream tests indicate that this simplification does not affect our results significantly. Scattering Transforms (see Section \ref{s:st}) require this continuity, and have advantages when the signal length is a power of two. 4096 is the largest power of two that fits comfortably inside the typical continuous TESS segment without including large gaps. About 95 percent of the light curves pass these criteria. The adopted length represents a compromise between signal coverage and continuity.  

Each accepted light curve is detrended as follows. First we compute a 5-sigma clipped mean flux in 100 equal-duration time bins. This provides a coarse estimate of the long-term trend while suppressing flares, eclipses, and other large excursions. We then fit and divide out a linear function through these binned means. The result is a robust first order correction and normalization that removes slow drifts with minimal distortion of astrophysical signals of interest.

We quantify the signal to noise ratio (SNR) of every detrended curve using three metrics. The standard deviation of the full curve (which all have a mean of 1) defines $SNR$. The point to point quantity $\mathrm{SNR}_{\mathrm{p2p}}$ measures the variance of consecutive differences, which for most sources is dominated by instrumental noise, on these timescales that are shorter than the typical intrinsic variability. We use the prescription of  \citet{nardielo20}. We further define 
$\mathrm{SNR}_{\mathrm{var}} = \frac{\mathrm{SNR}}{\mathrm{SNR}_{\mathrm{p2p}}}$, a measure of the amplitude of the overall variability relative to the noise floor of that light curve.

As discussed in Section \ref{s:model_select}, and in more detail in the appendix, we tested several additional normalization schemes beyond the mean flux of unity obtained when detrending. These included dividing by the standard deviation, by the median absolute deviation, by $\mathrm{SNR}_{\mathrm{p2p}}$, applying a hyperbolic tangent transformation to the fluxes, and combinations thereof. Some of these improved downstream performance of certain models, but not for our selected model. We therefore adopt the detrended curves with no further rescaling. 

Each final light curve is a 4096 element vector of (approximately) evenly sampled fluxes with mean fixed to one and first order trends removed. The following sections describe the transformations applied to these standardized curves to extract information.

\subsection{Periods and Metadata}

We obtain stellar and observational metadata for each target directly from the TESS archive, indexed by their TESS Input Identifiers (TIC; \citealt{stassun19}). For validation and controlled experimentation we use the variability catalog of \citet{fether}, which contains more than 70000 periodic variables identified in the first two years of the TESS Prime Mission. This catalog provides vetted periods, some coarse variability types, and quality metrics. In later sections we use subsets of this catalog for training and evaluation. Periods are used only for validation metrics and not in feature construction or as input to  training.

\section{Methods}
\label{s:methods}
We would like to map the set of about 1.5 million light curves in the 2-minute sample, to a continuous space where objects with similar light-curves cluster. While this space could have an arbitrarily large number of dimensions, it is impractical to interpret spaces with more than a few dimensions directly. It is reasonable to expect stellar light curves to not require more than a few dimensions to mostly capture their variance. In the following subsections we detail how we transform every 4096-element light curve and reduce its dimensionality, all the way to a single node assignment. We map all the light curves to a 2D hexagonal grid of size  $13 \times 13$ (169 nodes). We find that this is large enough to separate the light curves into a meaningful but not overwhelming number of assignments. 

Optimally, each node contains only objects that are similar to each other, and average properties change gradually and continuously across the map.

\subsection{Quantile Graphs}
\label{s:qg}

Quantile Graphs (QGs), introduced by \citet{campanharo11}, provide a conceptual duality between time series and networks or graphs. Here, following \citet{fsilva23} we use them as a coarse but efficient summary of the dynamical behavior of a light curve. The idea is intentionally minimalistic. The flux range of a light curve is discretized into a fixed and small number of quantiles. The quantiles are defined over the sample (see details in Section \ref{s:model_select}). Each quantile represents a node in a directed graph. For a chosen step size $k$, we track how often the light curve moves from quantile $i$ to quantile $j$ over exactly $k$ time steps. These transition counts populate an adjacency matrix $A_{ij}^{(k)}$ for every $k$. Diagonal entries represent `self loops' where the flux remains in the same quantile over that interval. For a given number of quantiles $q$ and a set of step sizes ${k}$, the final representation is the concatenation of all such adjacency matrices. In Section \ref{s:model_select} we find better results when the matrices are not scaled. The resulting dimensionality of this representation is $q^2$ per value of $k$. 

The quantiles can be defined either per light curve or globally across the sample. In the global scheme used here, the quantile boundaries are computed once from a random subset of the training set and then fixed for all light curves. This means that transitions between quantiles represent comparable flux changes across the dataset, rather than relative changes within each individual light curve. For a given model with $q$ quantiles, we find the edges by sampling $10^3$ random light curves, which is about $4\cdot10^6$ fluxes. We found that increasing the sample size further had no significant effect. 

While crude, the QG structure compactly encodes a discrete set of time derivatives at multiple lags. QGs have several practical advantages. They require no learned parameters, no training phase, and no dependence on the global dataset (except for the quantile values and the inevitable hyper-parameters). The computation for a single light curve of length $n$ scales linearly with $n$ and runs in about a millisecond on a 2020 laptop. This makes QGs attractive when the goal is to map millions of light curves or more.

A further advantage is conceptual extensibility, which we will explore in future publications. The scheme implicitly assumes uniform sampling because transitions are indexed by integer step size. For non-uniform data one could replace “step size” with actual time differences and reweight transitions accordingly. This is conceptually similar to the method of \citet{mahabal11,mahabal12}. Both approaches emphasize local temporal increments rather than absolute flux values, and are therefore naturally insensitive to phase and global offsets. Also, a QG can be continuously updated and monitored for second order variability. This offers the potential to detect transient phenomena on top of secular variability, which is a difficult problem. Furthermore, studying the topology of the graphs may prove useful to understand the light curves \citep{campanharo11}.  

\subsection{Scattering Transform}
\label{s:st}

Scattering Transforms (STs) provide a description of temporal structure by combining wavelet decompositions with predefined nonlinearities \citep{mallat11}. An ST maps a one-dimensional signal into a hierarchy of wavelet modulus coefficients that capture variability on multiple timescales while suppressing sensitivity to phase and small shifts. 

We use standard first-order and second-order scattering coefficients derived from a Morlet wavelet bank using the Matlab implementation. We try combinations of scale parameters that span the full TESS window from the Nyquist limit up to roughly half the length of the sequence. The resulting representation typically contains a few thousand coefficients. STs are more computationally expensive than QGs, but remain tractable for our dataset, taking a few tens of milliseconds per light curve. As we show below, despite the larger dimensionality compared to QGs, they seem to only preserve marginally more shape information, but with a less ordered result. 

\subsection{PCA reduction}
\label{s:pca}

Both QGs and STs produce high-dimensional feature vectors. We apply Principal Component Analysis (PCA) to the full training set of representations and retain the top $N_{\mathrm{pc}}=200$ components. This compression step reduces noise, accelerates SOM training, and allows different representation to be compared on an equal footing. As described in Section \ref{s:model_select}, we scanned different values of $N_{\mathrm{pc}}$ and found that skipping PCA or keeping too few components degrades the organization of the SOM, while increasing the count beyond 200 yields diminishing returns. We train the PCA once during training, and only apply it during evaluation. 

\subsection{Self Organizing Maps}
\label{s:som}

Self Organizing Maps (SOMs) are used to project high-dimensional data onto a low-dimensional discrete manifold while preserving local topology \citep{kohonen95}. See similar applications in \citet{brett04,armstrong16}. Each SOM consists of a set of nodes arranged on a regular 2D grid. Each node has an associated weight vector in feature space. During training the SOM iteratively adjusts these weight vectors so that nearby nodes respond similarly to nearby data points. For our purposes a SOM serves as an atlas of light curve similarity. We wish for light curves that are close in representation space (preferably and predominantly due to their similar shape) to cluster in contiguous regions of the map, while divergent morphologies should occupy distant nodes. Following our parameter search in Section \ref{s:model_select}, we use a hexagonal grid of size $13 \times 13$ (169 nodes), 100 learning epochs, 100 cover steps, an initial neighborhood of 3, and the default learning schedule. 

SOMs, compared to the large number of available methods for dimensionality reduction and clustering,  provide a compromise between flexibility and structure. They define a fixed, discrete low-dimensional manifold with an explicit mapping from feature space, preserve local topology by construction, and allow new data to be embedded independently once training is complete. These properties are key for the applications considered in this work.

\subsection{Learned vs. fixed components}
\label{s:learned}

Although QGs and STs themselves are deterministic given a light curve and a choice of hyperparameters, the full pipeline includes learned components. The PCA projection and the SOM weights are optimized during training.

The capacity of these learned components, by which we mean their ability to represent arbitrarily complex or highly nonlinear structure, is intentionally limited. PCA provides a linear compression that retains only the dominant variance directions, while the SOM represents the data using a fixed, low-resolution grid with strong smoothness constraints. As a result, the mapping cannot memorize individual light curves or fit fine-scale noise, and instead emphasizes stable, hopefully global, light curve structure.

     \begin{figure}
         \centering
    	\includegraphics[width=\columnwidth]{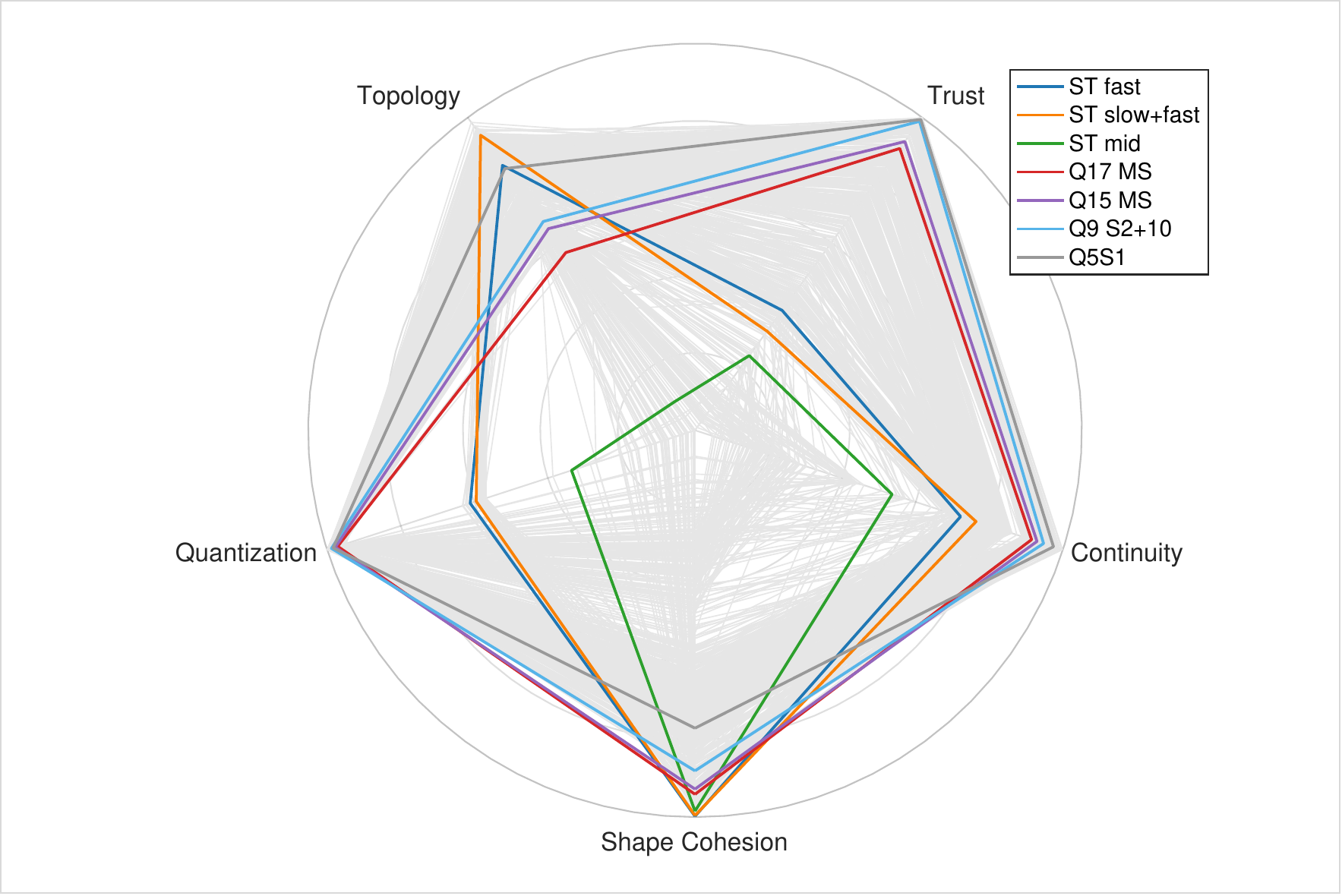}
     	\caption{Comparison of representative model configurations across five evaluation metrics: shape cohesion, quantization error, topological error, trustworthiness, and continuity. Metrics are rescaled so that better performance lies toward the perimeter. Gray polygons show all $\sim 1500$ evaluated configurations; colored curves highlight selected models discussed in the text. QG-based representations generally perform better than ST configurations on most metrics.}
        \label{f:radar}
    \end{figure}

     \begin{figure*}
         \centering
    	\includegraphics[width=\textwidth]{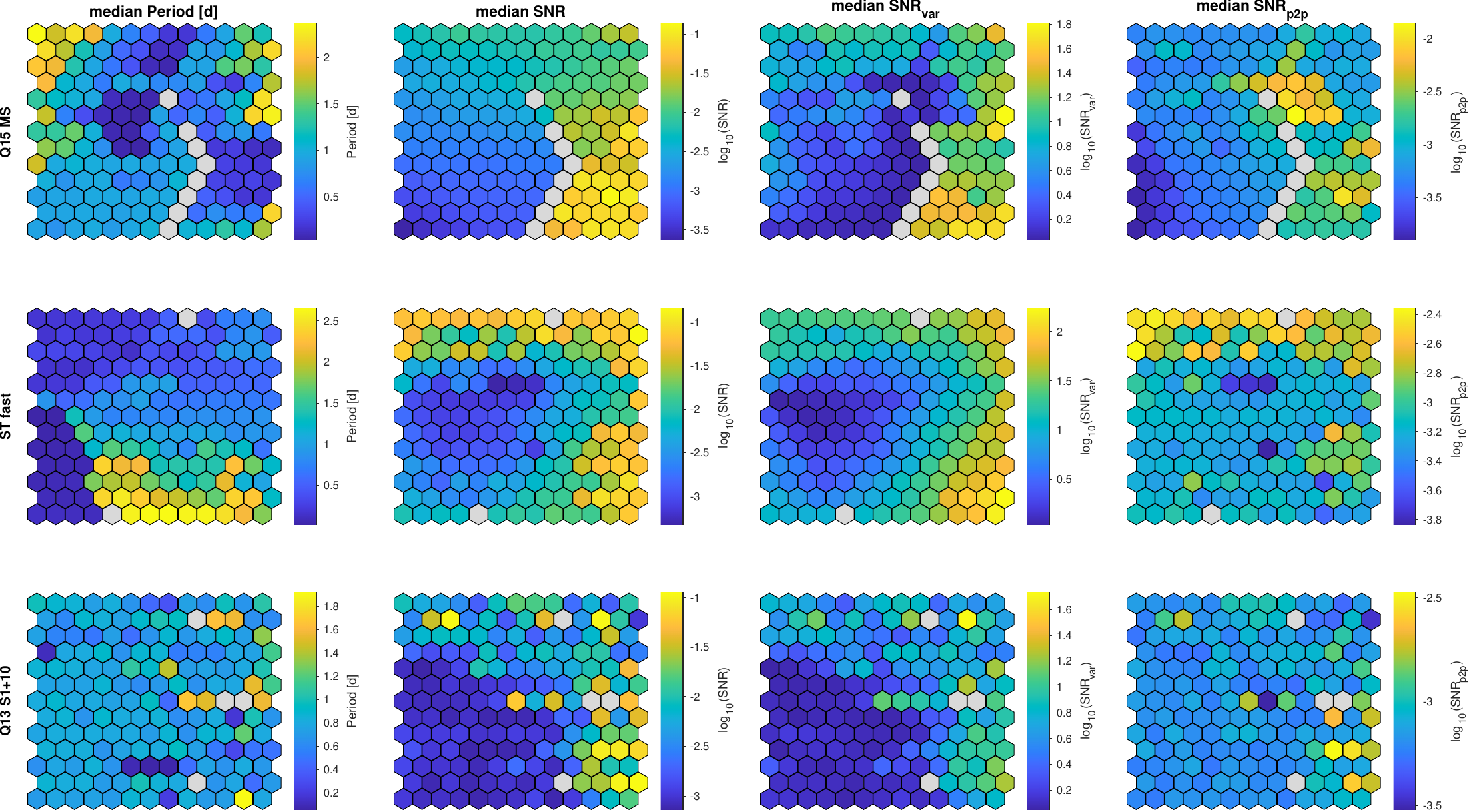}
     	\caption{Node-level statistics for 3 SOMs during our hyperparameter search, the best QG (top row), the best ST (middle row), and one of the worst models for contrast (bottom row). Clearly, the more successful SOMs trace the periodic signals, and correlate with the period as well as the various measures of variability or SNR.}
        \label{f:hexes}
    \end{figure*}

\section{model selection} \label{s:model_select}

Assigning a light curve to a node, its best matching unit (BMU), collapses thousands of
time samples into a single discrete index. The path from raw light curve to BMU
involves multiple design choices: detrending, optional normalization, the
representation family (QG or ST) and its hyperparameters, whether PCA reduction
is applied and at what rank, and finally the SOM geometry and details. 
Each of these choices can shift BMU assignments in nontrivial ways.

Although this is formally an unsupervised task, here we explicitly scan the
parameter space and evaluate models against a set of quantitative diagnostics,
focusing on SOM smoothness, robustness to noise, and morphological cohesion.
The goal is not to identify a provably optimal configuration but to select a
representation that is stable, interpretable, and useful across the relevant
variability regimes.

The most challenging criterion for model comparison is the morphological or \emph{Shape Cohesion} ---
whether neighboring light curves in a given mapping actually exhibit similar
shape. We approach this by first defining a cyclic cross-correlation phased-distance
(Appendix~\ref{a:circ}). For each model we compute, for every non-empty SOM
node, the median pairwise cyclic distance among all period-folded light curves assigned to
that node, and then average this statistic over all nodes. Models that group
together light curves of similar morphology yield lower cohesion scores.  
This metric evaluates periodic shape similarity and therefore favors organization consistent 
with periodic morphology, which is the dominant kind of variability in TESS. It is obviously not optimal for 
sources with more than one dominant period, or a stochastic behavior. 

To this we add four standard SOM quality metrics: the \emph{quantization
error} \citep{ultsch09}, the \emph{topological error} \citep{kiviluoto96}, 
the \emph{trustworthiness}, and the
\emph{continuity} \citep{venna05,venna07}. These diagnostics, widely used in assessing nonlinear
embeddings, capture distinct failure modes of low-dimensional mappings.

\paragraph{Quantization Error (QE).}
The quantization error measures how well the SOM represents the data locally.
For each light curve we compute the Euclidean distance between its feature
vector and the weight vector of its BMU. The QE is the average of these
distances over the full dataset. Lower QE indicates that the SOM nodes lie
closer, on average, to the data they represent.

\paragraph{Topological Error (TE).}
The topological error quantifies violations of neighborhood preservation. 
For each light curve we identify both its BMU and its
second-best matching unit. If the latter is not adjacent to the former on the
SOM grid, the mapping breaks local topology. The TE is the fraction of data
points for which this occurs. A perfectly topological map has TE = 0.

\paragraph{Trustworthiness.}
Trustworthiness (T) evaluates whether close neighbors in the low-dimensional
SOM space correspond to true neighbors in the original feature space. 
For each point, T penalizes cases where the SOM introduces
false neighbors that were not close in the high-dimensional representation.
Trustworthiness ranges from 0 to 1, with higher values indicating fewer such
artifacts. We use a neighborhood size of 10.

\paragraph{Continuity.}
Continuity (C) measures the converse failure mode: whether true neighbors in
the high-dimensional space remain neighbors after projection to the SOM. 
Low continuity indicates that the mapping tears apart clusters that are close in the original representation. Like trustworthiness, continuity ranges from 0 to 1, with higher values indicating better preservation of global structure.

Together these metrics provide a multi-faceted assessment of performance:
Shape cohesion probes intra-node morphological similarity; QE captures representation
accuracy; TE tests topological stability; and trustworthiness and continuity
probe neighborhood preservation in both directions. 

For model comparison, we build a training and a test set of about 14K light curves each, with periods shorter than 3d, from \citet{fether}. Since our light curves are 4096 time-steps long with a 2-minute cadence, they are under 6d long. Longer periods would not repeat even once. As detailed in the appendix, we tried different light curve normalization schemes, different pre-PCA normalizations, multiple SOM sizes and parameters, and various combinations of QGs and STs. For every set of hyperparameters we generate a SOM and measure its score on these 5 metrics on the test set. We ran about 1500 fits, varying combinations of these parameters, using different models, as detailed in the appendix.

Figure \ref{f:radar} provides a qualitative comparison of the different model configurations explored in our search. For visualization purposes, the five evaluation metrics are standardized and rescaled so that better performance lies toward the perimeter of the diagram. Each polygon therefore summarizes the relative performance of a given configuration across all diagnostics. The background gray polygons represent the full set of evaluated configurations, while the colored curves highlight representative models that achieve relatively good Shape Cohesion. In general, QG representations outperform ST configurations on most metrics, although the best ST models achieve slightly better shape-cohesion values.

In Figure \ref{f:hexes} we show a few per-node statistics for 3 SOMs, our selected QG model (see Section \ref{s:best}), the best ST of similar size (13x13), and one of the worst SOMs for contrast. Note how The top two SOMs with small errors are much smoother, with adjacent nodes often having similar values. 
\\

\section{Selected Model}\label{s:best}
We select a QG model with 15 quantiles, $q=15$, times 8 skip values, $\{k\}=\{1,4,8,16,32,64,128,256\}$. Though the choice among the top models is somewhat subjective, it is not arbitrary. There is no single way to combine our 5 metrics, but this model performed relatively well on all of them, while also being the least sensitive to other pipeline choices. Given that we next train on a larger dataset, and infer on a sample even larger and more diverse, this seems like the most robust choice. 

We do not normalize the light curves beyond detrending, and reduce the dimensionality of the representation from 1800 to 200 principal dimensions with PCA. We then train a SOM, with an hexagonal geometry of size $13 \times 13$, using Manhattan distances, 100 cover steps, 100 epochs of training, and an initial neighborhood of 3. We now train this model again, using the entirety of the $P<3d$ sample from \citet{fether}, over 46,000 light curves. 

The resulting SOM has three empty nodes ($\#9,23,161$), a singleton ($\#36$), and 165 nodes with more than 10 light curves. The median node count is 230, and the biggest node ($\#65$) clusters over 1300 light curves. As seen during model selection, the map correlates relatively smoothly with our various measures of SNR and variability, with the most variable objects generally populating the bottom left of the SOM, and the noisiest/least variable stars near the top and right edges of the map. While the nodes are not unimodal, in any sense of the word, they do generally cluster similar-looking light curves. In order to extract a prototypical light curve from every node, we use the same logic used for the cyclic phased distance, to align, stack, and find a median phased light curve as well as the MAD of the training set, in every populated node. We refer to these templates as medoids. They capture the shape of the dominant periodic signal in a node.

\begin{figure*}[p]
\centering
\includegraphics[width=0.9\textwidth]{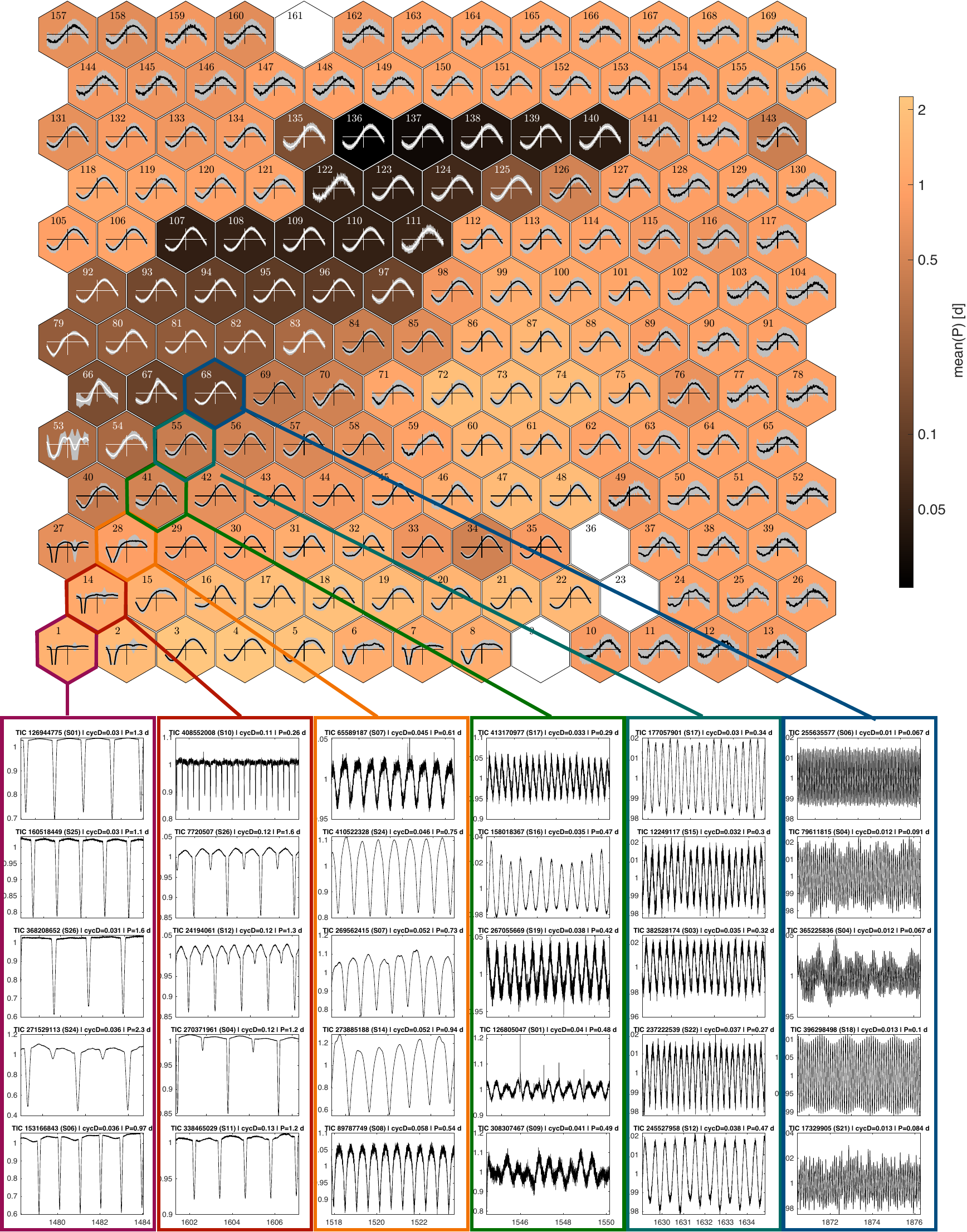}

\caption{We map our main training set of 46K light curves to our fiducial SOM. 
Every node in the top panel is colored according to the median period of the training set in the node. 
We superpose in black the median-aligned, scaled, and stacked light curves in every node, 
with the scatter ($\pm2\,\mathrm{MAD}$) shown in gray. 
Below, we follow a progression of nodes from the bottom left toward the middle, 
showing the five light curves closest to the BMU. 
The SOM organizes light curves according to a combination of SNR, amplitude, period, and signal shape.}

\label{f:mega}
\end{figure*}

 The top panel of Figure \ref{f:mega} shows the SOM, with nodes colored by their mean (log) period, and the scaled medoid light curve superimposed. The gray band around every medoid is at $\pm2\mathrm{MAD}$. The SOM is highly correlated with the period, as can be seen from the continuous shading in the figure, but also with the SNR, as is apparent when looking at the scatter around the medoid. The scatter is markedly higher nearer the top and right edges of the map, compared to the bottom left, for example, but the progression is mostly gradual. From the narrow scatter around many medoids (especially at higher SNR) it is clear that the SOM also captures shape efficiently. 
For example, in the bottom panel of Figure \ref{f:mega} we start from node \#1 at the bottom left of the SOM, and follow a `Top-Right' progression through nodes \#14,28,41,55, and 68. For each node we show 5 of the best fitting light curves in the training set, ranked by cyclic distance. These nodes are quite representative of what we generally see in the SOM. While the nodes are clearly clustering by period and SNR, they also display rather similar light-curve morphologies, though it is far from perfect. 

In the most minimal sense, what we have arrived at is a pipeline that assigns a node to any given TESS light curve. This prediction, which includes the calculation of the adjacency matrices, PCA reduction, and then transforming to the SOM plane, takes about a millisecond per source on a 2020 laptop. In the following section we apply it to the full 2-minute cadence catalog of about 1.5M light curves. 

\section{Application to the full sample}
\label{s:2mn}

We apply the trained SOM to the full TESS 2-minute cadence catalog that passes our selection criteria, comprising nearly 1.5M light curves from over 500K unique TICs. About 50K light curves do not pass these criteria and are not included. We emphasize that we do not select for variability or SNR in any way, nor do we use periods, labels, or any auxiliary information beyond the light curves themselves. All light curves are embedded independently and assigned to their best-matching node on the 13$\times$13 SOM. 

The full sample populates 167 of the 169 nodes, one more than the training set. Node 161, which was empty during training, contains only three objects in the full catalog. The median node occupancy is 3185 light curves, but node sizes vary widely. The largest node (node 156) contains over 150K light curves, which largely show little or no detectable intrinsic variability. 

\begin{figure*}
\centering
\includegraphics[width=\textwidth]{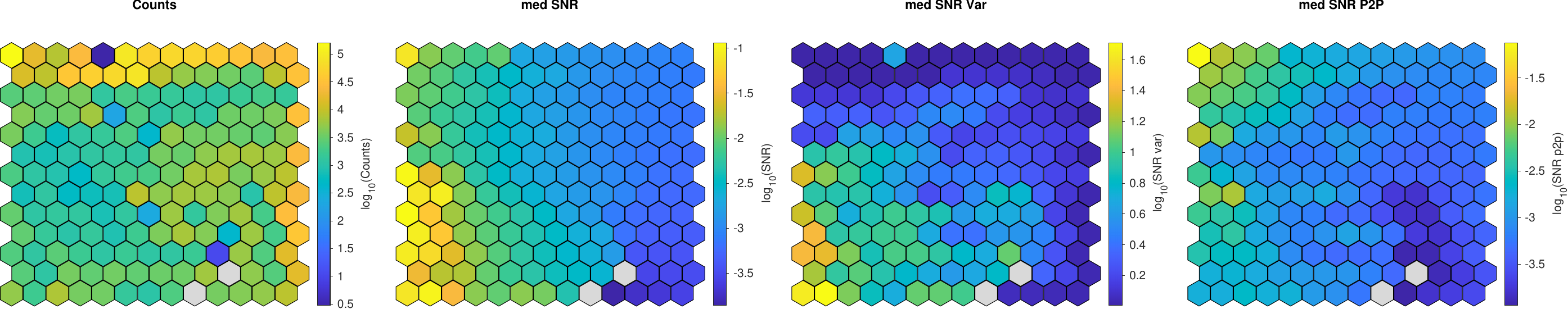}
\caption{Node-level statistics for the full 2-minute catalog. The SOM correlates strongly with various measures of variability amplitude and SNR, in close analogy to the training set (Figure~\ref{f:hexes}). A large number of low-variability light curves aggregate near the edges of the map.}
\label{f:2m_hexes}
\end{figure*}

As shown in Figure~\ref{f:2m_hexes}, the SOM maintains a structure similar to that seen for the small training set during the hyper-parameter search (top row of Figure~\ref{f:hexes}), with node-level averages of SNR and variability amplitude varying smoothly across the map. A large and expected fraction of light curves concentrate in a small number of nodes near the edges of the SOM, which are dominated by low-SNR and low-variability sources.

In addition to a node assignment, we calculate for each light curve its ranked distance from the BMU, denoted $D_{\rm bmu}$. 
The distances in representation space within each node (in the evaluation set) are ranked and converted to percentiles. Objects with small $D_{\rm bmu}$
lie near the node core, while values approaching 100 indicate increasingly distant members.

To assess whether this quantity, while noisy, still provides a useful indicator of membership quality, we performed a blinded calibration experiment. We generated approximately 240 sets of 24 light curves drawn from 40 randomly selected nodes. Within each node, light curves were grouped by similar $D_{\rm bmu}$ values and presented to the author alongside the eight closest light curves to the BMU. Each set was scored qualitatively by estimating how many of the 24 appeared visually consistent with the node core.

\begin{figure}
\centering
\includegraphics[width=\columnwidth]{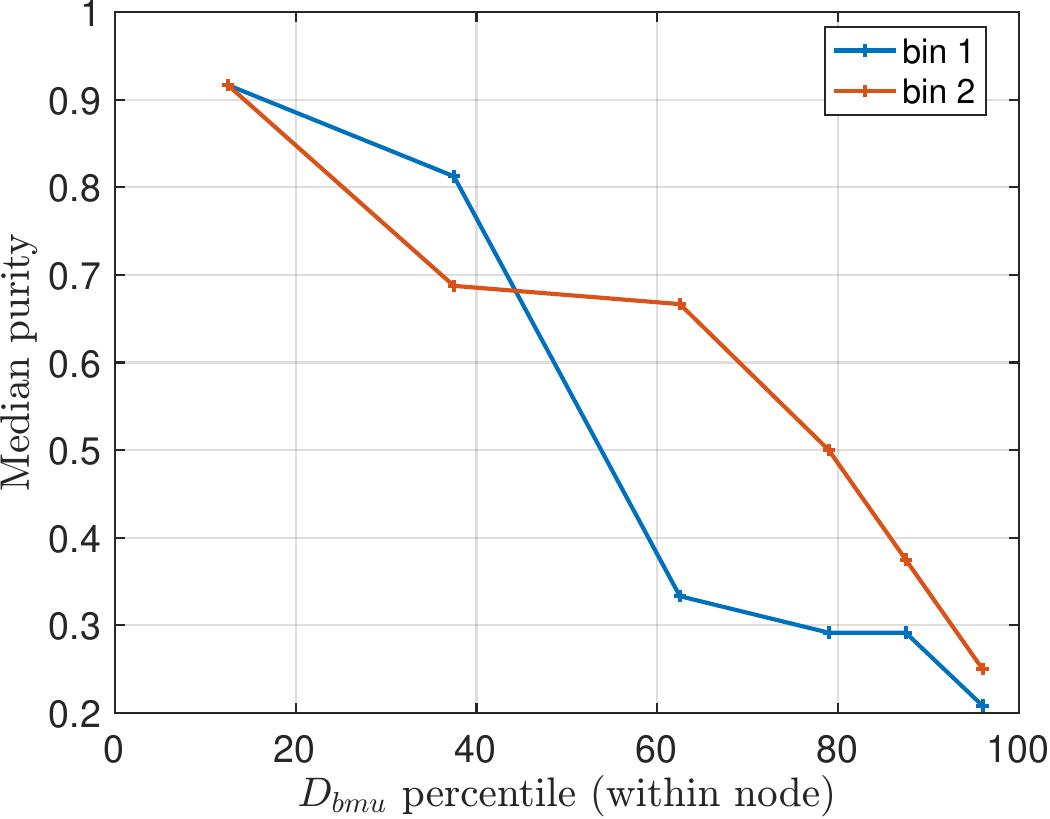}
\caption{Purity (estimated by eye) as a function of percentile-ranked distance from the node center, in bins of SNR. On average, $D_{\rm bmu}$ provides a useful indicator of confidence in node membership.}
\label{f:pur_bmu}
\end{figure}

Despite the coarse and subjective nature of this purity measure, we find that $D_{\rm bmu}$, is a reasonable indicator of membership quality within a node. As shown in Figure~\ref{f:pur_bmu}, when results are binned by SNR, the inferred purity measure starts above $\sim$90\% at small $D_{\rm bmu}$ percentiles and declines steadily to $\sim$25\% near the 95th percentile, independent of SNR. 

We further find that a large majority of the tested nodes (86\%) exhibit a negative Spearman rank correlation between purity and $D_{\rm bmu}$, with a median correlation coefficient of $\rho_{\rm med} = -0.85$. Nodes where this behavior breaks down are predominantly low-SNR. In practice, this implies that the ranked distance can usually be interpreted as a measure of confidence in node membership, except in the noisiest regions of the map. Typical objects have small rank values, while outliers and misclassified light curves become dominant for ranks above $\sim$90\%.

\subsection{Repeat observations}\label{s:repeats}

Approximately half of the stars in the sample were observed in more than one sector, with a median of three sectors and a long tail extending to a few hundred stars which were observed in over 40 sectors. For each star we define $n_{\rm sector}$ as the number of observed sectors and $n_{\rm node}$ as the number of distinct SOM nodes to which its light curves are assigned.

\begin{figure}
\centering
\includegraphics[width=\columnwidth]{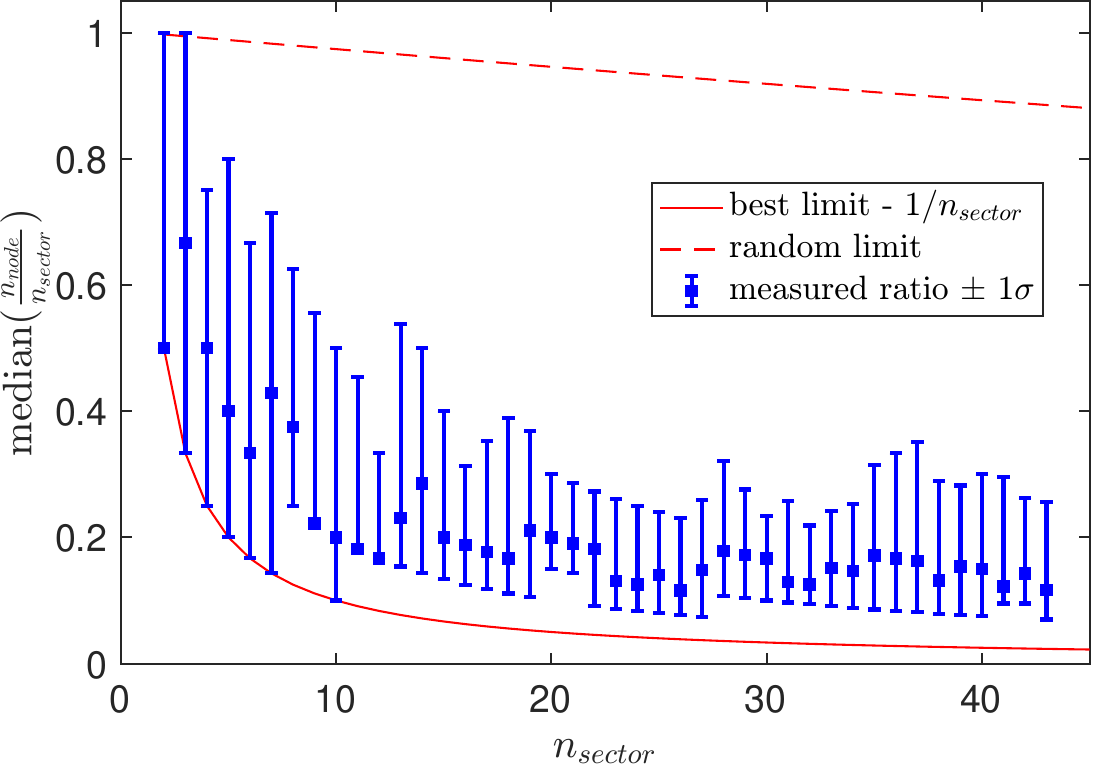}
\caption{We count how often repeat observations of the same star ended up in different nodes. If assignments were random the number of nodes would be close to the number of independent observations and the (blue) measurements would be near the dashed red line. Instead we are much closer to the red line, the optimal limit were all the sources are assumed to be persistent and noiseless and assignments are perfectly stable.}
\label{f:repeats}
\end{figure}

In the idealized noiseless limit, persistent sources would always map to the same node, yielding $n_{\rm node}=1$. Conversely, if node assignment was dominated by noise or intrinsic variance, $n_{\rm node}$ would grow approximately linearly with $n_{\rm sector}$. As shown in Figure~\ref{f:repeats}, the observed behavior lies much closer to the idealized limit than to the random expectation, indicating that repeated observations typically occupy a small, stable neighborhood of the map.

This repeat-based behavior provides a natural mechanism for anomaly detection. Stars that exhibit transient or state-dependent variability on top of an otherwise persistent signal will move across the SOM between sectors. By identifying stars whose sector-to-sector node assignments span unusually large distances on the map, or whose assignments depart from their typical neighborhood in a single sector, it becomes possible to flag candidate transient phenomena or state changes for further inspection. 

The SOM is not perfectly consistent and continuous geometrically, such that sometimes objects with similar light curves populate nodes that could be further apart, mostly due to differences in SNR or period, and the limitations of representing everything in two dimensions. We therefore expect Euclidean distances on the SOM to be noisy tracers of state change. To investigate this, we constructed an empirical neighborhood graph based on repeat observations. For each pair of SOM nodes, we counted how often light curves from the same star, observed in different sectors, were assigned to both nodes, and normalized these counts by node occupancy. This defines a symmetric, data-driven similarity graph on the SOM nodes that is independent of the map topology. Starting from this graph, we generated an animation in which nodes are initially shown as distinct regions and progressively merged as the similarity threshold is lowered. The nodes were rendered on the SOM grid with colors encoding their spatial position, allowing the merging process to be inspected visually.

\begin{figure}
\centering
\includegraphics[width=\columnwidth]{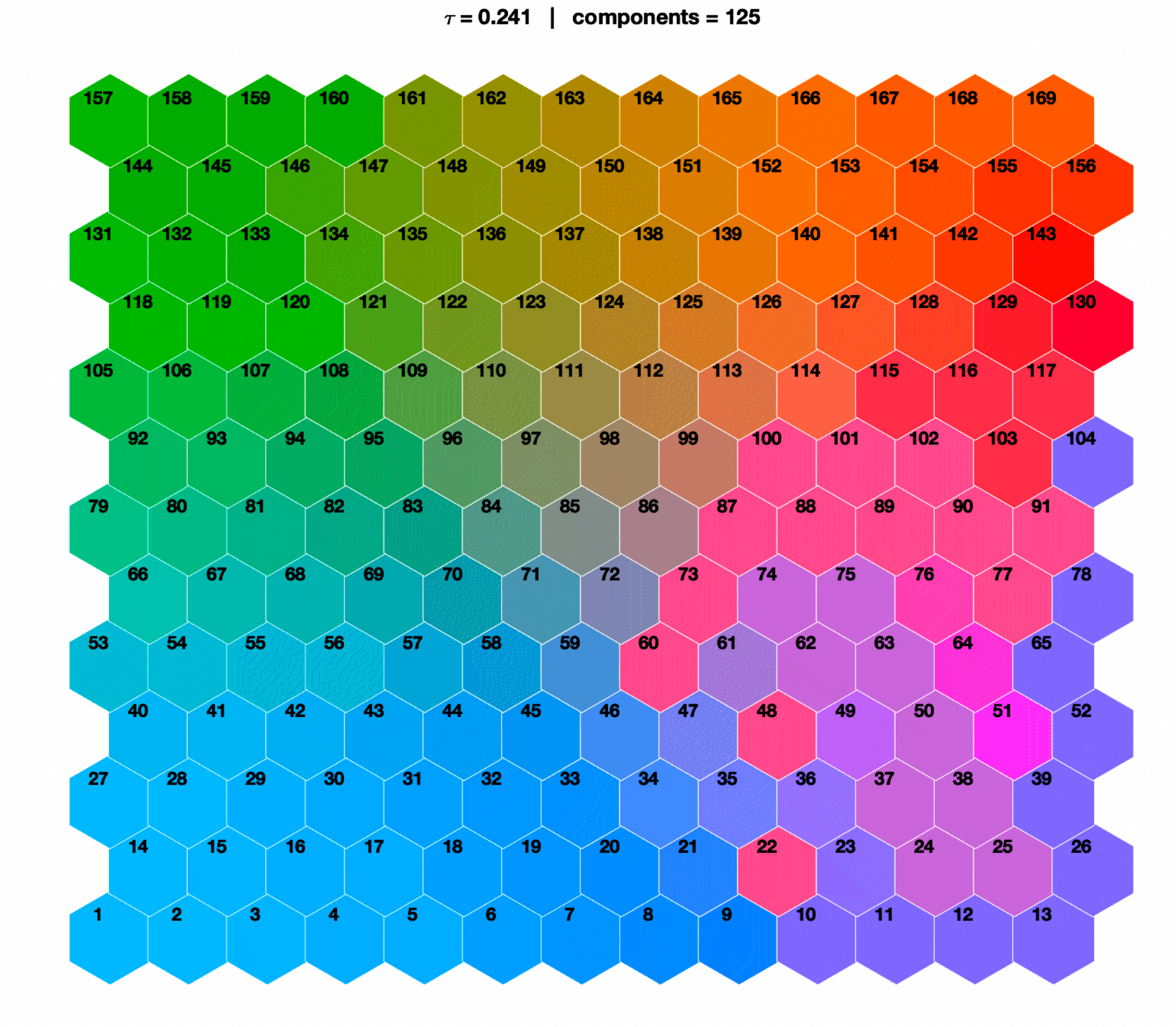}
\caption{Intermediate frame from the repeat-based node-merging animation. Colors indicate merged regions at a fixed similarity threshold. Most regions are contiguous on the SOM grid, indicating strong consistency between the learned topology and empirical neighborhood relations derived from repeat observations.}
\label{f:gif}
\end{figure}

The resulting animation reveals that, over a wide range of similarity thresholds, merged regions are overwhelmingly contiguous on the SOM grid, indicating that the learned topology is largely consistent with empirical neighborhood relations derived from repeated observations. We show a representative frame of this animation in Figure \ref{f:gif}. Only a small number of non-local merges are observed, and these tend to be coherent and persistent across thresholds, suggesting genuine non-convex structure in the learned representation rather than noise-driven artifacts. This qualitative agreement between the SOM geometry and the repeat-based neighborhood graph provides additional validation of the embedding, while also highlighting its limitations. In particular, it motivates the use of graph-based distances on the learned neighborhood structure, rather than simple Euclidean distances on the SOM grid, when comparing nodes or identifying outliers.

Distances between light curves can be defined at multiple stages of the pipeline, including in the original QG feature space, in the PCA-reduced space, and on the SOM grid itself. These distances are not equivalent and should be interpreted with care. Euclidean distances on the two-dimensional SOM grid provide only a coarse proxy for similarity. Because the SOM necessarily distorts the high-dimensional geometry, equal grid distances do not correspond to equal dissimilarities in representation space. This effect is particularly pronounced across regions where variability amplitude or signal-to-noise dominates the embedding. The repeat-observation analysis here provides an empirical, data-driven alternative. By constructing a similarity graph based on how often light curves from the same star co-occupy pairs of nodes across sectors, we obtain a notion of neighborhood that is independent of the SOM geometry. Graph-based distances derived from this structure are therefore better motivated for comparing nodes or identifying outliers.

\section{Interactive web tool}
\label{s:webtool}

We provide an interactive web-based interface to explore the learned map and its associated TESS light curves. \href{http://tess-l8.space}{tess-l8} is designed to give immediate, visual insight into the relationship between individual light curves, their assigned SOM nodes, and the global structure of the map. In practice, the tool enables several use cases, including the identification of outliers or rare phenomena, and the inspection of individual objects in the context of their nearest neighbors. 

At the highest level, the interface presents the full 2-min sample distributed over the hexagonal SOM grid (Figure~\ref{f:web_som}). Each cell corresponds to a node, and can be colored by various summary statistics (e.g., median variability signal-to-noise, amplitude proxies, or node-level quality diagnostics). This enables rapid identification of regions of interest, such as high-variability clusters or sparsely populated nodes. Selecting a node opens a detailed view showing basic statistics of the light curves assigned to the node, the medoid derived from the training set and its dispersion, and a tunable number of light curves. One can also download a table with the full population of a node. 

The tool also supports inspection at the level of individual objects. Given a TIC identifier, the interface aggregates all available sectors and displays their corresponding SOM assignments. For each sector, the user is shown the detrended (thinned to conserve disk space) light curve, the assigned node, and basic diagnostics such as distance to the BMU, variability metrics, and rank within the node. This allows users to assess the stability of the mapping across sectors and to identify cases where the same object is assigned to different regions of the map, as can be seen in the example in Figure \ref{f:web_tic}. A detailed view for a single TIC and sector additionally shows a set of nearest neighbors to that light curve from the training set within the node, ranked by proximity in feature space. We further provide a direct link to search the star on \href{http://simbad.cds.unistra.fr/simbad/}{SIMBAD} \citep{wenger00}. For example, following these links for stars in  node $\#36$ reveals that they are mostly known Asymptotic Giant Branch (AGB) stars.

\section{Discussion and Conclusions}
\label{s:conclusions}

\subsection{What the representation captures}

Across the analyses presented above, a consistent picture emerges of what aspects of the light curves dominate the embedding.

First, the representation strongly organizes light curves by variability amplitude and signal-to-noise ratio. This is evident from the smooth gradients across the SOM, and from the concentration of low-variability sources in specific regions of the map. These quantities define much of the large-scale structure, and is consistent across the different metrics we explored. 

Second, despite not using periods as inputs, the map is highly correlated with period for periodic sources. This indicates that the representation captures temporal structure closely related to periodicity, even without explicit folding. 

Third, within regions of comparable SNR and period, the SOM further separates light curves by morphology. Nodes typically contain families of visually similar light curves, and the medoids provide meaningful summaries of these shapes.

Finally, repeat observations show that the representation is stable under independent measurements of the same source. Most stars map to a small and contiguous region of the SOM across sectors, indicating that the embedding captures persistent properties rather than noise realizations.

Taken together, the representation primarily encodes a combination of variability amplitude, characteristic timescale, and light-curve shape, with noise acting as a secondary driver.

\subsection{Use cases}

The representation is not designed to directly produce scientific results without further analysis. Instead, it provides an intermediate layer that supports interaction with large light-curve datasets.

One primary use case is exploratory analysis. By organizing millions of light curves into a structured map, the representation provides a global overview of variability in the survey. This can guide targeted follow-up, inform the construction of training sets, and highlight regimes of variability that are underrepresented or poorly understood.

A second use case is quality control. Because the mapping is fast and deterministic once trained, it can be applied uniformly to large samples and to newly ingested data. Abrupt changes in node occupancy, unusual concentrations of sources, or systematic shifts across observing sectors can indicate instrumental artifacts or pipeline issues.

A third use case is anomaly detection. Outliers within a node can be identified using ranked distances, while discrepancies between node assignment and metadata can reveal unusual objects. Repeat observations provide an additional signal: sources that move across the map between sectors may indicate transient or state-dependent behavior.

Finally, the representation can serve as a front-end to downstream methods. Node assignments and low-dimensional embeddings provide compact inputs that can be combined with metadata or domain-specific features in supervised or semi-supervised analyses.

\subsection{Limitations and failure modes}

Paraphrasing Tolstoy, every representation is biased in its own way \citep{tolstoy1877}. Our mapping reflects the design choices made in preprocessing, feature construction, and model selection, and should not be interpreted as a complete or neutral description of variability in TESS.

Variability amplitude and signal-to-noise ratio exert a strong influence on the embedding. In low-SNR regimes, morphology becomes poorly constrained and node assignments are less stable.

Although periods are not used as inputs, the representation was developed and selected using periodic sources over a limited range of periods, and remains correlated with period for known periodic variables. This introduces a bias toward periodic structure.

The mapping is also sensitive to preprocessing choices, including detrending, normalization, and window selection. While we explored several alternatives, the space of possible choices is large, and different applications may favor different configurations.

Finally, compressing millions of light curves into a two-dimensional manifold inevitably introduces distortions. Non-convex structure in the high-dimensional space may be folded or stretched, leading to cases where similar light curves appear in non-adjacent nodes.

Despite these limitations, the representation is simple, fast, and scalable. It has been applied to the full 2-minute cadence catalog and can be extended to larger datasets with minimal additional cost. Its value lies in providing a stable and interpretable organization of the data that reduces the complexity of subsequent analysis.

\section*{Acknowledgments}
I  thank L. Bouma, S. Shahaf, M. J. Graham, L. Hillenbrand, and T. Prince for insightful thoughts and help at various stages. 
I acknowledge support from Israel Science Foundation (ISF) grant 541/17, and by grant 2018017 from the United States-Israel Binational Science Foundation (BSF).
This research was also funded in part by the Koret Foundation, the Kavli Institute for Particle Astrophysics and Cosmology at Stanford University, and by grant NSF PHY-2309135 to the Kavli Institute for Theoretical Physics (KITP). 

This paper is based primarily on data collected with the TESS mission, obtained from the MAST data archive at the Space Telescope Science Institute (STScI). Funding for the TESS mission is provided by the NASA Explorer Program. STScI is operated by the Association of Universities for Research in Astronomy, Inc., under NASA contract NAS 5–26555. This research used resources of the National Energy Research Scientific Computing Center (NERSC), a Department of Energy User Facility.

\bibliographystyle{apsrev4-1}

\bibliography{main}

\clearpage

 \begin{appendix}

\section{Method details}
\subsection{Shape Cohesion}\label{a:circ}
We require a distance measure between two phase–folded light curve templates
that is insensitive to circular phase shifts. For a reference template \(a\) and
one or more comparison templates \(b_i\), we compute a \emph{circular
cross-correlation distance} as follows.

Each template is first normalized to unit Euclidean norm so that the comparison
depends only on shape and not amplitude. For every \(b_i\) we then compute its
full circular cross-correlation with \(a\) using FFTs, which efficiently
returns the correlation for all possible phase lags. For each \(b_i\), we
identify the lag that maximizes the real part of the cross-correlation; this is
taken as the optimal circular phase alignment between the two curves. Let
\(C_i\) denote the magnitude of the cross-correlation at this best-matching
lag. When two shapes match perfectly up to a shift, \(C_i = 1\); when they are
orthogonal, \(C_i = 0\).

We convert this similarity measure into a distance via
\[
D_i = \frac{1}{C_i} - 1 .
\]
This mapping satisfies \(D_i = 0\) for identical shapes (up to phase), and
\(D_i\) increases monotonically as the match worsens. This provides a fast, phase-invariant measure of shape similarity between periodic light curves.

\section{Hyperparameter Search and Model Selection}
\label{app:hpsearch}

Although the representation learning pipeline is mostly unsupervised, it contains a number of design choices that affect the resulting embedding, as well as a few learned components. We therefore performed a structured hyperparameter search over a finite grid of model configurations. The grid was defined over ranges that we determined in preliminary testing. The goal was not to exhaustively optimize every detail, but to identify a configuration that yields stable topology, low distortion, and reasonable intra-node cohesion, while remaining computationally tractable at survey scale.

We varied parameters at three levels: (i) light-curve normalization, (ii) representation family and feature preprocessing, and (iii) SOM properties. The explored ranges are summarized in Table~\ref{t:hpspace}. The total number of configurations we explored is not the Cartesian product of the above choices, which would have been too computationally onerous. Each training and evaluation requires over a minute, and this product is in the tens of thousands. We did however run over 1500 combinations of the best performing ranges of each parameter, keeping the others fixed near their optimum. 

\begin{table*}
\centering
\begin{threeparttable}
\caption{Hyperparameter search space explored in this work.}
\label{t:hpspace}
\begin{tabular}{lll}
\hline
Category & Parameter & Explored Values \\
\hline
Light-curve preprocessing 
& Normalization method 
& none, p2p, mad,p2p+mad, tanh+mad, zscore$^a$ \\
\hline
Quantile Graph (QG) 
& Number of quantiles $q$ 
& 7, 9, 11, 12, 13, 15, 17$^b$\\
& Lag set $\{k\}$ 
& $\{1\}$, $\{10\}$, $\{1,10\}$, $\{2,10\}$, $\{1,4,8,16,32,64,128,256\}^b$ \\
& Quantile definition
& Global (i.e., defined over the sample)$^c$\\
\hline
Scattering Transform (ST) 
& Implementation 
& MATLAB \\
&  Quality Factors $[Q_1,Q_2]$
& $[2,1]$, $[8,1]$, $[16,1]$,  concatenation of $[2,1]$ and $[16,1]$\\
& Invariance Scale
& Signal length (4096)$^d$\\
\hline
Principle Component Analysis (PCA)
& Normalization method
& none, per\_feature, per\_block \\
& Number of components 
& 200 (fixed after preliminary tests) \\
\hline
Self Organizing Map (SOM)  
& Grid size $m\times m$ 
& $11\times11$, $13\times13$, $15\times15^b$ \\
& Cover steps 
& 50, 100 \\
& Initial neighborhood 
& $\lfloor f\,m \rfloor$, $f\in\{0.25,0.5,0.75\}$ \\
& Training epochs 
& 100, 200 \\
& Geometry
& Hexagonal\\
& Distance metric 
& Manhattan$^b$ \\
\hline
\end{tabular}

\begin{tablenotes}
\item[a] 
mad denotes division by the median absolute deviation; 
p2p denotes scaling by the standard deviation of point-to-point differences 
(consecutive flux differences); 
tanh+mad applies a hyperbolic tangent transformation followed by mad scaling; 
zscore denotes per-light-curve mean subtraction and division by the standard deviation.
\item[b] Additional values were explored but not as systematically.
\item[c] When defining the quantiles per object, we find the best performance with QGs is obtained with zscore normalization. This is equivalent to the simpler scheme of no normalization and a global quantile definition, which we opt for. 
\item[d] Initial testing showed significant degradation with smaller values.
\end{tablenotes}

\end{threeparttable}
\end{table*}

For each configuration, we trained the PCA and SOM on $14,000$ periodic $P<3$\,d  light curves from \citet{fether}, and evaluated performance on a held-out test subset of the same size. We derive the five diagnostics defined in Section \ref{s:model_select}: cyclic shape cohesion ($\mathrm{cyc\_D}$), quantization error ($Q_{\mathrm{err}}$), topological error ($T_{\mathrm{err}}$), trustworthiness, and continuity. Primary emphasis was placed on minimizing $\mathrm{cyc\_D}$, preferring models that performed consistently well across all metrics. Table~\ref{t:hpresults} lists the results for our selected model, the best performing configuration by each metric, a model with bad SOM organization for comparison, and a few randomly selected examples.

\begin{sidewaystable*}
\centering
\scriptsize
\setlength{\tabcolsep}{3pt}
\begin{threeparttable}
\caption{Representative model performance on the held-out test set.}
\label{t:hpresults}

\begin{tabular}{lllc|ccccc|l}
\toprule
\multicolumn{4}{c|}{Input} & \multicolumn{5}{c|}{Evaluation} & Notes \\
\cmidrule(lr){1-4}\cmidrule(lr){5-9}\cmidrule(lr){10-10}
Model & Hyper-parameters & Norm (LC/PCA) & SOM $(m,f,c_s,e)$
& $\mathrm{cyc\_D}$ & $Q_{\mathrm{err}}$ & $T_{\mathrm{err}}$ & Trust & Cont & \\
\midrule

Q15 MS & q=15, k=\{1,4,8,16,32,64,128,256\} & none/none & (13,0.25,100,100) & 0.1328 & 1.247 & 0.1218 & 0.9553 & 0.9708 & Fiducial model \\
ST fast & Q=[16,1] & p2p/none & (13,0.25,100,100) & 0.07893 & 4.285 & 0.1899 & 0.7819 & 0.908 & Best $\mathrm{cyc\_D}$ \\
Q7 S2+10 & q=7, $k=\{2,10\}$ & p2p/none & (15,0.5,100,100) & 0.2023 & 1.154 & 0.2697 & 0.9908 & 0.9814 & Best Trust \\
Q7 S2+10 & q=7, $k=\{2,10\}$ & zscore/none & (15,0.75,100,100) & 0.181 & 1.175 & 0.1244 & 0.9902 & 0.987 & Best Continuity \\
ST fast & Q=[16,1] & tanh-mad/per\_block & (15,0.25,100,100) & 0.2079 & 1.033 & 0.1848 & 0.8931 & 0.9285 & Best Quantization (Qerr) \\
Q11 MS & q=11, k=\{1,4,8,16,32,64,128,256\} & none/none & (11,0.75,100,100) & 0.1417 & 1.18 & 0.0652 & 0.9557 & 0.9759 & Best Topology (Terr) \\
Q13 S1+10 & q=13, k=\{1\} & p2p-mad/per\_block & (13,0.25,100,100) & 0.3194 & 1.431 & 0.34 & 0.7177 & 0.7129 &  Model with bad SOM organization \\
\midrule
Sample range & \multicolumn{3}{l|}{mean $\pm$ std over all evaluated configurations} & 0.239$\pm$0.094 & 1.68$\pm$1.5 & 0.221$\pm$0.082 & 0.881$\pm$0.11 & 0.911$\pm$0.081 & \\
\midrule
Q9 S2+10 & q=9, $k=\{2,10\}$ & mad/none & (13,0.77,100,100) & 0.1974 & 1.258 & 0.1851 & 0.9878 & 0.9812 & Randomly selected example \\
Q15 S1+10 & q=15, $k=\{2,10\}$ & zscore/per\_block & (11,0.55,100,100) & 0.3033 & 1.431 & 0.2901 & 0.6988 & 0.7642 & Randomly selected example \\
ST mid & Q=[8,1] & p2p/none & (15,0.27,100,100) & 0.08457 & 4.106 & 0.2382 & 0.8269 & 0.9149 & Randomly selected example \\
Q17 MS & q=17,  k=\{1,4,8,16,32,64,128,256\} & p2p-mad/none & (13,0.77,100,100) & 0.1807 & 1.129 & 0.182 & 0.9768 & 0.9642 & Randomly selected example \\
Q12 S1+10 & q=12, $k=\{1,10\}$ & p2p-mad/none & (11,0.27,100,100) & 0.1515 & 1.184 & 0.2911 & 0.9834 & 0.9704 & Randomly selected example \\



\bottomrule
\end{tabular}

\begin{tablenotes}
\item Lower values are better for $\mathrm{cyc\_D}$, $Q_{\mathrm{err}}$, and $T_{\mathrm{err}}$. Higher values are better for trustworthiness and continuity.
\item
The SOM hyperparameters are shown as $(m,f,c_s,e)$ where $m\times m$ is the grid size,
$f$ is the initial neighborhood factor ($\mathrm{init\_neigh}=\lfloor f\,m\rfloor$),
$c_s$ is the number of cover steps, and $e$ is the number of epochs.
\end{tablenotes}
\end{threeparttable}
\end{sidewaystable*}

The adopted model corresponds to a QG representation with $q=15$, the multi-scale lag set described above, $N_{\rm pc}=200$, and a $13\times13$ SOM trained with 100 cover steps and 100 epochs. While several similar configurations yield comparable performance, this model performed well for most hyper-parameters we tested. Since Our eventual training set is a few times larger than during the parameter search, and our target population is more diverse than the training set, this model seemed least likely to fail. We emphasize that the purpose of this search was not to identify a unique global optimum, but to ensure that the reported results are reasonably general and stable.

\section{Additional figures}
\label{ap:figs}
\begin{figure*}[t]
    \centering
    \includegraphics[width=0.9\linewidth]{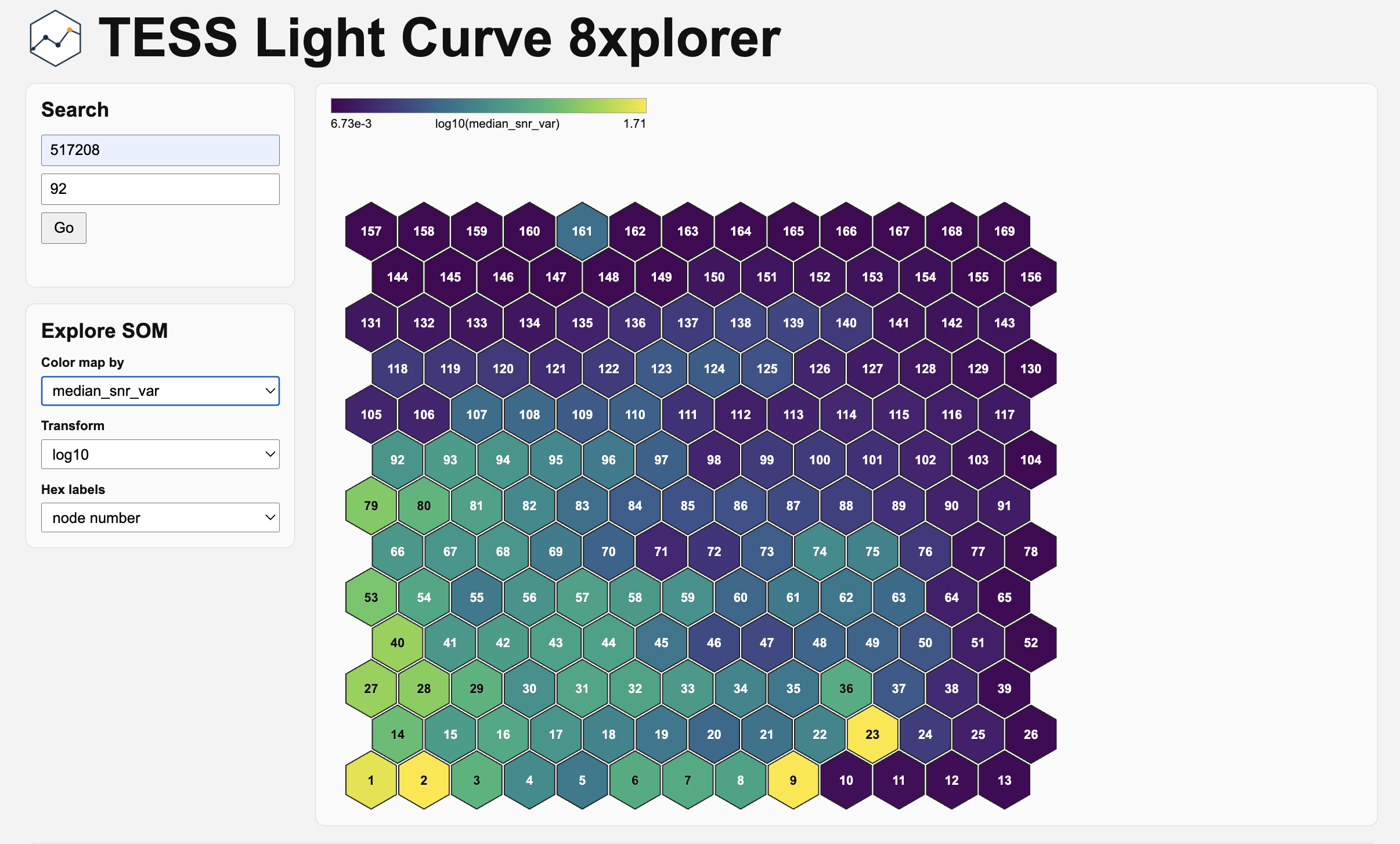}
    \caption{Overview of the SOM in the \href{http://tess-l8.space}{web interface}. Each hexagonal cell represents a node, colored here by a variability proxy. The map provides a global view of the organization of light-curves in the learned feature space, and an entry point to the catalog. Clicking on a node brings node-level information and a view of the light curves associated with it.}
    \label{f:web_som}
\end{figure*}

\begin{figure*}[t]
    \centering
    \includegraphics[width=0.9\linewidth]{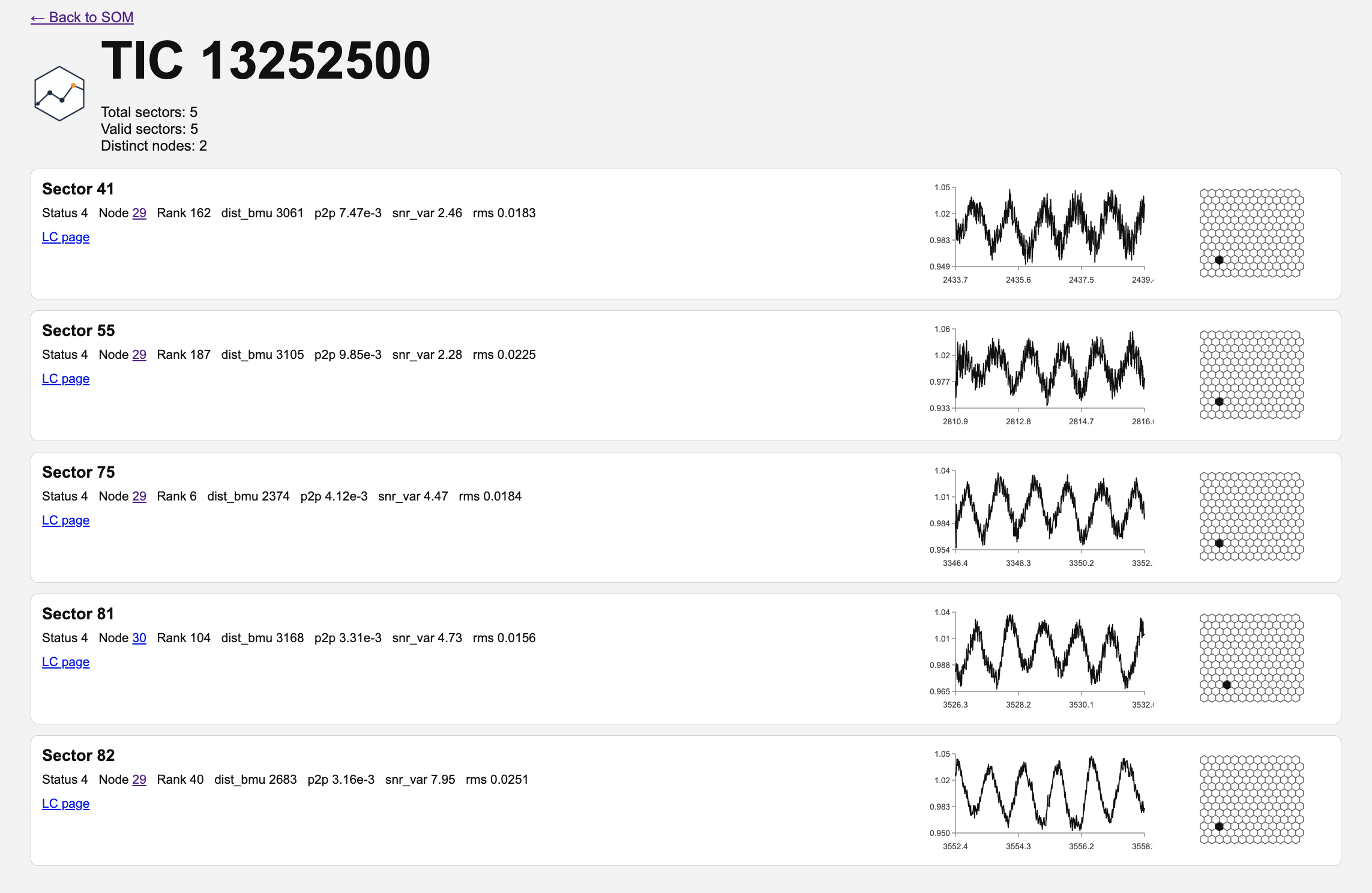}
    \caption{Example TIC-level view in the \href{http://tess-l8.space}{web interface}. The page displays all the sectors where this star was observed and where they are mapped on the SOM. This allows to quickly identify objects that change state, or nodes that have similar sources. }
    \label{f:web_tic}
\end{figure*}

We show here screen captures from \href{http://tess-l8.space}{tess-l8}, the web interface to our model. 
Figure~\ref{f:web_som} Shows the top level view of the SOM, where every node is clickable and leads to the relevant node page. Figure \ref{f:web_tic} shows an example view of a given star, indexed by TIC, with the sectors it was observed on. This allows users to assess the stability of the mapping across sectors and to identify cases where the same object is assigned to different regions of the map.

 \end{appendix}

\end{document}